\documentstyle[preprint,eqsecnum,aps,epsfig]{revtex}
\tightenlines

\begin{document}
\draft
\preprint{HUPD-0103}

\title{Dynamical Fermion Masses Under the Influence of Kaluza-Klein 
Fermions in Randall-Sundrum Background
\footnote{This paper is dedicated to the 60th birthday of 
Professor Hagen Kleinert who is the good old friend of one of 
the authors. }}

\author{Hiroyuki Abe$^1$, Tomohiro Inagaki$^2$, and Taizo Muta$^3$\\\ }

\address{$^1$ Department of Physics, Hiroshima University,\\
Higashi-Hiroshima, Hiroshima 739-8526, Japan\\
E-mail: abe@theo.phys.sci.hiroshima-u.ac.jp\\\ }
\address{$^2$ Research Institute for Information Science and Education, 
Hiroshima University,\\
Higashi-Hiroshima, Hiroshima 739-8521, Japan\\
E-mail: inagaki@hiroshima-u.ac.jp\\\ }
\address{$^3$ Department of Physics, Hiroshima University,\\
Higashi-Hiroshima, Hiroshima 739-8526, Japan\\
E-mail: muta@hiroshima-u.ac.jp}

\maketitle

\begin{abstract}
The dynamical fermion mass generation on the D3-brane in the 
Randall-Sundrum space-time is discussed in a model with 
bulk fermions in interaction with fermions on the branes. 
It is found that the dynamical fermion masses are generated at 
the natural (R.-S.) radius of the compactified extra space 
and may be made small compared with masses of the Kaluza-Klein modes
which is of order of TeV.
\end{abstract}

\narrowtext

\section{Introduction}

\noindent
It is an interesting idea to assume an existence of the 
extra-dimensional space which eventually compactifies 
leaving our 4-dimensional space-time as a real world.\cite{KK} 
The recent proposal\cite{Ant,Ark1} for the mass scale of the compactified 
space to be much smaller than the Planck scale gave a strong 
impact on the onset of studying phenomenological evidences of 
extra-dimensional effects.\cite{Han} 
There is a crucial problem, however, how such large extra dimensions 
are stabilized. Recently, 
L. Randall and R. Sundrum\cite{Ran} (R.-S.) gave an alternative 
to the large extra dimension scenario. By introducing a specific curved 
bulk space they succeeded to have a mass scale much smaller than the Planck 
scale without relying on the fine-tuning. 

In our present analysis we introduce bulk fermions 
in the R.-S. space-time\cite{Cha} and see what effects could be observed.
The bulk fermions interact with 
themselves as well as with fermions on the 4-dimensional branes 
through the exchange of the graviton and its Kaluza-Klein excited 
modes, or through the exchange of gauge bosons which may be 
assumed to exist in the bulk.\cite{Cha} The interactions among 
fermions generated as a result of the exchange of all the Kaluza-Klein 
excited modes of the graviton or gauge bosons may be expressed 
as effective four-fermion interactions.\cite{Han,Dob,Chen} 
According to the four-fermion interactions we expect that 
the dynamical generation of fermion masses will take place.
In the present communication we look for a possibility of the dynamical 
fermion mass generation under the influence of the bulk fermions through 
the effective four-fermion interactions in the R.-S. background.
In the R.-S. background fermion mass terms are forbidden 
by the $S^1/{\bf Z}_2$ symmetry. 
The possible source of 
fermion masses on the branes is two-fold, i. e. the dynamically 
generated fermion masses and masses of the Kaluza-Klein excited modes
of the bulk fermions. The mass of the Kaluza-Klein excited modes is
known to be of order of TeV.\cite{Cha} 

In section \ref{sec2} we will review the dynamical fermion mass generation with 
bulk fermions in {\it torus} compactified case\cite{Abe1} because most of the 
technical parts can be summarized in the torus (flat) extra dimension case.
And in section \ref{sec3} we proceed to the case of the {R.-S.} space-time, the main 
part of this paper. In the section 
it is found that the dynamical fermion masses are generated at 
the natural (R.-S.) radius of the compactified extra dimensional space 
and may be made small compared with masses of the Kaluza-Klein modes
which is of order of TeV because of the presence of 
the Randall-Sundrum warp factor.

\section{Flat Extra Space}\label{sec2}
\noindent
We assume an existence of 5-dimensional bulk fermions $\psi$ 
in interaction with fermions $L$ on the 4-dimentional brane.
Effective interactions among these fermions can be given in the form 
of the four-fermion interaction. 
In fact it is known that the exchange 
of the Kaluza-Klein excited modes of the bulk graviton results in 
effective four-fermion interactions.\cite{Han} 
After the Fierz transformation on the four-fermion interactions 
we generate the transition-type interactions.
Accordingly we start with the following Lagrangian for our model\cite{Abe1}
\begin{equation}
   {\cal L\/}^{(5)}  
         =  \bar{\psi} i \gamma^M \partial_M \psi  \, + \, [ \,
             \bar{L} i \gamma^\mu \partial_\mu L 
              \, + \, g^2 \bar{\psi} \gamma^M L  \, 
                        \bar{L} \gamma_M \psi \, ] \, \delta( x^4 ),    
\label{eq:sLag}
\end{equation}
where $g$ is the coupling constant with mass dimension -3/2 and index $M$
runs from 0 to 4 while index $\mu$ runs from 0 to 3. 
Fermions $\psi$ and $L$ are assumed to be of $N_f$ components.

After neglecting tensor interactions and introducing auxiliary field 
$\sigma \sim \bar{\psi}L$, we compactify the bulk space on torus 
with radius $R$. Our 4-dimensional Lagrangian 
is rewritten as 
\begin{eqnarray}
  {\cal L\/}^{(4)} 
      =   \bar{\Psi} (M + i /\!\!\!\partial) \Psi 
                  \ - \  \left| \sigma \right|^2 ,
\end{eqnarray}
where 
$\Psi^t \equiv 
       \left( L \, , \, \psi_{0} \, , \, \psi_{1} \, , \,
           \psi_{-1} \, , \, \psi_{2} \, , \, \psi_{-2} 
           \cdots
        \right)$
and 
\begin{eqnarray}
  M \equiv      
       \left(
         \matrix{
            0 & m^{*} & m^{*} & m^{*} & m^{*} & m^{*} & \cdots \cr
            m & 0     & 0     & 0     & 0     & 0     & \cdots \cr
            m & 0     & \frac{1}{R}   & 0     & 0     & 0 & \cdots   \cr
            m & 0     & 0     & - \frac{1}{R} & 0     & 0 & \cdots   \cr
            m & 0     & 0     & 0     & \frac{2}{R}   & 0 & \cdots   \cr
            m & 0     & 0     & 0     & 0     & - \frac{2}{R} & \cdots   \cr
            \vdots & \vdots & \vdots & \vdots & \vdots & \vdots & \ddots  
         }
       \right), \qquad (m=g\sigma/\sqrt{2\pi R}). \label{M} 
\end{eqnarray}
If $\sigma$ acquires a non-vanishing vcuum expectation value,
we replace $\sigma$ in $m$ by its vacuum expectation value 
$\left\langle \sigma \right\rangle$, 
i. e. $m = N g \left\langle \sigma \right\rangle$. 
The eigenvalues of matrix $M$ determine the masses of 4-dimensional fermions. 
Obviously we find 
that the lightest eigenvalue is given by
\begin{eqnarray}
\lambda_{\pm 0} = \pm \ |m| \quad {\textrm{for}} \quad |m| \ll 1/R. 
\label{Mass}
\end{eqnarray}
Thus we conclude that within our scheme there is a possibility of having the 
light fermion masses which is much smaller than the mass of the Kaluza-Klein
modes of the bulk fermion. 
By performing the path-integration for fermion field $\Psi$ 
we find the effective potential for $\sigma$ 
in the leading order of the $1/N_f$ expansion:
\begin{eqnarray}
V(\sigma)  =  |\sigma|^2 &-& \frac{1}{2 \pi^2}
                \int_0^\Lambda dx \, x^3
                \ln \left[x^2 + |m|^2 (\pi xR) \coth(\pi x R)\right] 
                \nonumber \\
          &-& \frac{1}{2 \pi^2} \sum_{j=1}^{\infty}\, 
                \int_0^\Lambda dx \, x^3
                \ln \left[x^2 + \left( \frac{j}{R} \right)^2\right].
\end{eqnarray}
The gap equation to determine the vacuum expectation value 
$\left<|\sigma|\right>$ of $|\sigma|$ reads
\begin{eqnarray}
\frac{\partial V(\sigma)}{\partial |\sigma|} 
   =  2 |\sigma| \, \left\{ 1- \frac{g^2}{2 \pi^2}\,
      \int_0^\Lambda dx \, \frac{x^3}{2 x \tanh(\pi x R) + g^2 |\sigma|^2}
      \right\} = 0.
\label{Gap}
\end{eqnarray}
By numerical observation of Eq. (\ref{Gap}) we find that there exists 
a non-trivial solution for $|\sigma|$ for a suitable range of parameters 
$g$ and $R$ and the solution corresponds to the true minimum of the 
effective potential. Accordingly the fermion mass is generated dynamically. 
Here the auxiliary field $\sigma$ (or the composite field $\bar{L}\psi$) 
acquires a vacuum expectation value.
Moreover it is easily confirmed that 
the phase transition associated with this symmetry breaking is of 
second order.

\begin{figure}[htbp]
   \centerline{\epsfig{figure=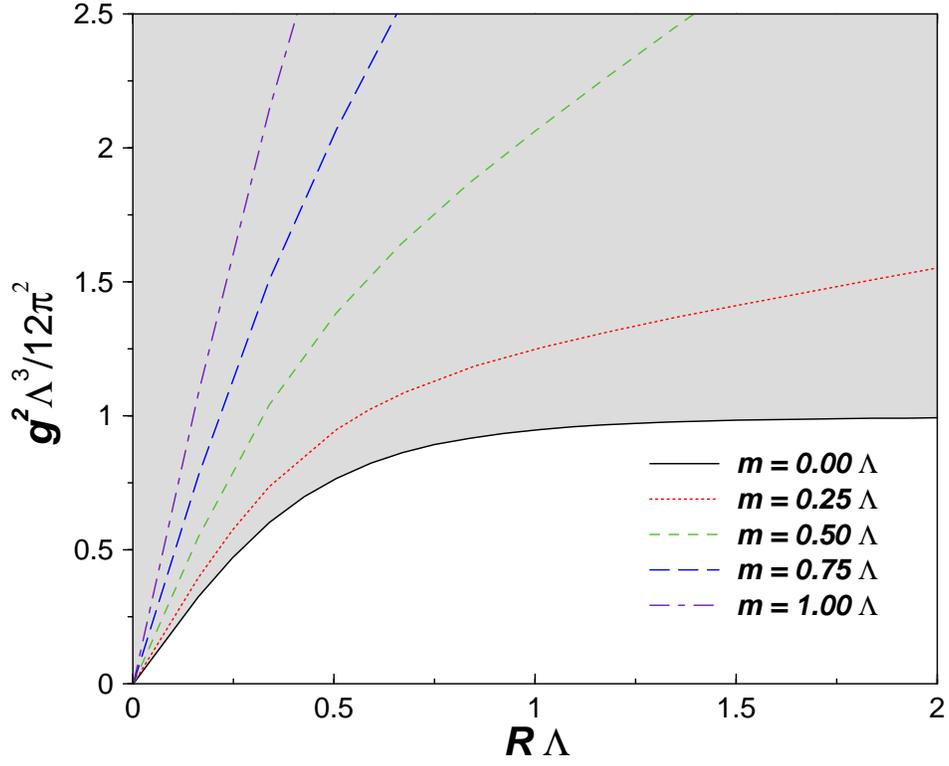,width=0.75\linewidth}}
   \caption{Critical radius as a function of $g$.}
   \label{fig:g-R}
\end{figure}

As was shown in Eq. (\ref{Mass}) the lowest fermion mass on the 
4-dimensional brane is $m=Ng \left<|\sigma|\right>$ 
where $\left<|\sigma|\right>$ is 
determined by solving Eq. (\ref{Gap}). 
The critical curve which represents the critical radius as a function of the 
coupling constant is shown in Fig. \ref{fig:g-R} as the curve for $m=0$. 
If we assume $1/R \sim$ TeV, light ($<$ TeV) fermions are obtained 
in the region between 
the solid line and the dot-dashed line in the Fig. \ref{fig:g-R} 
because it is natural that $R\Lambda \sim 1$.
We, however, have no idea to justify $1/R \sim$ TeV and so we need to introduce 
a certain mechanism such as the Randall-Sundrum model in the next
section.

\section{Warped Extra Space}\label{sec3}
\begin{figure}[htbp]
\centerline{
\epsfig{figure=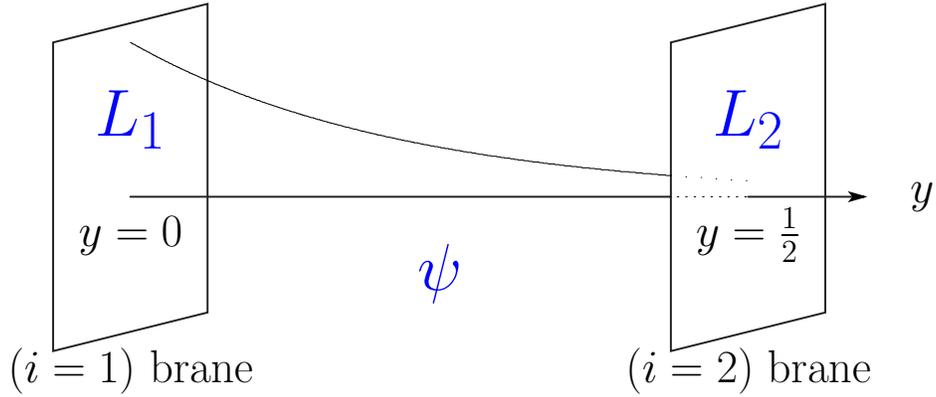,width=0.75\linewidth}}
\caption{Four-fermion interaction model in R.-S. background.}
\label{RSfig}
\end{figure}

\noindent
In theories with large ($\gg M_{pl}^{-1}$) 
extra dimemsions people wonder 
how such large radii are stabilized. 
We have not yet come across with any satisfactory 
answer, while important results in many models depend essentially on 
the largeness of the radius as in the previous section.
L. Randall and R. Sundrum\cite{Ran} noticed that the existence of 
the brane leads to the curved bulk space and proposed the 
so-called Randall-Sundrum (R.-S.) model.
As is shown in the Fig. \ref{RSfig} 
their model has {\it two} D3-branes on the $S_1/{\bf Z_2}$ orbifold fixed points 
and the {\it AdS$_5$} between these branes:
\begin{eqnarray}
ds^2 \equiv G_{MN} dx^M dx^N 
     = e^{-2kb_0|y|}\eta_{\mu\nu}dx^\mu dx^\nu + b_0^2dy^2, 
\end{eqnarray}
where $k \sim M_{pl}$ is the gravity scale, $b_0^{-1} \sim k/24\pi$ 
is the compactification scale. One of the most important results 
is the $e^{-\frac{1}{2}kb_0}$ ({\it warp factor}) suppression of the K.-K. 
masses of the bulk fields.\cite{Cha}

\subsection{Bulk Fermions in R.-S. Background}
\noindent
Following Chang {\it et al.}\cite{Cha} we derive the mode expansion 
of the bulk fermion in the R.-S. space-time such that 
\begin{eqnarray}
\psi(x,y) = 
  \frac{e^{\frac{3}{2}kb_0|y|}}{\sqrt{b_0}}
  \sum_n \left[ \psi_L^{(n)}(x){\xi (y)} 
              + \psi_R^{(n)}(x){\eta(y)} \right],
\label{modeex}
\end{eqnarray}
\begin{eqnarray}
\left\{
\matrix{
{\xi (y)}
  = \sqrt{\frac{kb_0}{1-e^{-\frac{1}{2}kb_0}}}
    e^{-\frac{1}{2}kb_0|\frac{1}{2}-y|}
    \sin \displaystyle{ \frac{m_n}{kb_0} }
    (1-e^{kb_0|y|}) \cr \cr
{\eta(y)} 
  = \sqrt{\frac{kb_0}{1-e^{-\frac{1}{2}kb_0}}}
    e^{-\frac{1}{2}kb_0|\frac{1}{2}-y|}
    \cos \displaystyle{ \frac{m_n}{kb_0} }
    (1-e^{kb_0|y|}) 
} 
\right.,
\end{eqnarray}
where 
$m_n \equiv {n\pi kb_0}/{(e^{\frac{1}{2}kb_0}-1)}$
is $b_0$ times the mass of the n-th K.-K. mode.
With this expansion the kinetic terms of the K.-K. modes become 
\begin{eqnarray}
{\cal L\/}^{(4)}_{\rm bluk} 
  = \int dy\, E \bar{\psi} i \gamma^M D_M \psi
  = \sum_{n} \left[ \bar{\psi}^{(n)}i /\!\!\!\partial \psi^{(n)}
      - \frac{m_n}{b_0} \bar{\psi}^{(n)} \psi^{(n)} \right].
\end{eqnarray}
In the following we apply previous results of our model with bulk fermions 
in the R.-S. background.

\subsection{Four-Fermion Interaction Model in R.-S. Background}
\noindent
Now we apply the previous torus-compactified model (\ref{eq:sLag})
to the case in the R.-S. background: 
\begin{eqnarray}
E_{\bar{M}}^{\ \ M}
      =  \left(
         \matrix{
          e^{kb_0|y|}\eta_{\bar{\mu}}^{\ \mu} & 0 \cr
               0      & b_0^{-1} 
         }
         \right),
\end{eqnarray}
where $E_{\bar{M}}^{\ \ M}$ is the vielvein whose square 
becomes the metric $G_{MN}$. We introduce the bulk fermion 
$\psi$ which propagates in the whole five dimensional space as 
in the torus case, while 
{\it two} brane fermions $L_i\ (i=1,2)$ propagate on an i-th brane 
shown in the Fig. \ref{RSfig} respectively.

We start with the Lagrangian in 5-dimensions,
\begin{eqnarray}
{\cal L\/}^{(5)}
     &=&  E \bar{\psi} i \gamma^M D_M \psi 
      +   E_{(1)} \left[ {
          \bar{L_1} i \gamma^\mu \partial_\mu L_1 
          + \frac{g_1^2}{N_f} \bar{\psi} \gamma^M L_1 
                    \bar{L_1} \gamma_M \psi \, } \right]
          \delta( x^4 )  \nonumber \\ && \hspace{1cm}
      +   E_{(2)} \left[ {
          \bar{L_2} i \gamma^\mu \partial_\mu L_2 
          + \frac{g_2^2}{N_f} \bar{\psi} \gamma^M L_2 
                    \bar{L_2} \gamma_M \psi \, } \right]
          \delta( x^4 \hspace{-0.3cm}-\textstyle{\frac{1}{2}}), 
\label{eq:5Lag}
\end{eqnarray}
where $x^M=(x^\mu,x^4=y),\ \mu=(0,1,2,3)$, 
$g_i$'s are the four-fermion couplings $([g_i]=-3/2)$ and 
$N_f$ is the number of the component of $\psi$ and $L_i$'s.
Note that the bulk mass term is removed by the $S^1/{\bf Z_2}$ projection.
Because Lorentz covariance should be preserved, 
we neglect the tensor (vector) interactions and obtain 
\begin{eqnarray}
{\cal L\/}^{(5)}
    &=&    E \bar{\psi} i \gamma^M D_M \psi
     +    \left[
          \bar{L_1} i \gamma^{\bar{\mu}} \partial_{\bar{\mu}} L_1 
          + \frac{g_1^2}{N_f} \bar{\psi} \gamma_5 L_1 
                    \bar{L_1} \gamma_5 \psi  \right]
          \delta( x^4 ) \nonumber \\ && \hspace{0.5cm}
     +    e^{-2kb_0}\left[ e^{\frac{1}{2}kb_0}{
          \bar{L_2} i \gamma^{\bar{\mu}} \partial_{\bar{\mu}} L_2 }
          + \frac{g_2^2}{N_f} \bar{\psi} \gamma_5 L_2 
                    \bar{L_2} \gamma_5 \psi  \right]
          \delta( x^4 \hspace{-0.3cm}-\textstyle{\frac{1}{2}}).
\end{eqnarray}
After chiral rotation $\psi \to e^{i \frac{\pi}{4} \gamma^5} \psi$ 
and $L_i \to e^{i \frac{\pi}{4} \gamma^5}L_i$, 
we introduce auxiliary fields $\sigma_i$'s to obtain 
\begin{eqnarray} 
   {\cal L\/}^{(5)}
       &=&  E \bar{\psi} i \gamma^M \partial_M \psi
         +  \left[ \bar{L_1} i /\!\!\!\partial L_1 
         -  N_f \left| \sigma_1 \right|^2 
         +  (g_1 \sigma_1 \bar{\psi}  L_1  +  h.c.) \right] 
            \delta( x^4 ) \nonumber \\ && 
         +  e^{-2kb_0}\left[ e^{\frac{1}{2}kb_0} 
            \bar{L_2} i /\!\!\!\partial L_2 
         -  N_f \left| \sigma_2 \right|^2 
         +  (g_2 \sigma_2 \bar{\psi}  L_2  +  h.c.) \right] 
            \delta( x^4 \hspace{-0.3cm}-\textstyle{\frac{1}{2}}).
\end{eqnarray}
Now we substitute the mode expansion of $\psi$ of Eq. (\ref{modeex}) 
into the Lagrangian and integrate it over the extra-dimension. Taking 
$e^{-\frac{3}{4}kb_0}L_2 \to L_2$ and $e^{-kb_0}\sigma_2 \to \sigma_2$,
Lagrangian in 4-dimension becomes 
\begin{eqnarray}
  {\cal L\/}^{(4)} 
&\equiv& \int dy \,{\cal L\/}^{(5)} \nonumber \\ 
&=& \sum_{n} \left[ \bar{\psi}^{(n)}i /\!\!\!\partial \psi^{(n)}
 -  \frac{n}{R} \bar{\psi}^{(n)} \psi^{(n)} \right] 
 +  \bar{L_1} i /\!\!\!\partial L_1 
 +  \bar{L_2} i /\!\!\!\partial L_2 
 -  N_f (\left| \sigma_1 \right|^2 + \left| \sigma_2 \right|^2) 
    \nonumber \\ &&
 +  \sum_n \left( \mu_1 \bar{\psi}_R^{(n)} L_1     
 +  h.c.\right) 
 +  \sum_n \left( (-1)^n \mu_2 \bar{\psi}_R^{(n)} L_2     
 +  h.c.\right),
\label{4LagRS}
\end{eqnarray}
where
\begin{eqnarray}
\mu_1 \equiv \displaystyle{\frac{g_1 \sigma_1}{\sqrt{b_0}}},\ \ 
\mu_2 \equiv \displaystyle{\frac{g_2 \sigma_2}{\sqrt{b_0}}},\ \ 
    R   \equiv \frac{b_0}{m_1} = \frac{e^{\frac{1}{2}kb_0}-1}{\pi k} 
         \sim    ({\rm TeV})^{-1}.
\end{eqnarray}
Here we note that the mass of the 1st K.-K. excited mode $1/R$ 
is of order of TeV without fine-tuning because of the warp factor 
in the R.-S. metric, and it is the only dependence on the factor.

After integrating out all fermionic degrees of freedom, 
we obtain the effective potential for $\sigma_i (\mu_i)$ as follows:
\begin{eqnarray}
\lefteqn{\bar{V}({\mu_1}/\Lambda,{\mu_2}/\Lambda)} 
\hspace{0cm} \nonumber \\
&=& [ \,V({\mu_1}/\Lambda,
            {\mu_2}/\Lambda) - V(0,0)\,] / \Lambda^4 
     \nonumber \\
&=& 2\pi R\Lambda
    \left[ \frac{({\mu_1}/\Lambda)^2}{g_1^2\Lambda^3} + 
           \frac{({\mu_2}/\Lambda)^2}{g_2^2\Lambda^3} 
           \right] \nonumber \\ &&
   -\frac{1}{2\pi^2} \int_0^1 dz\, z^3 
     \ln \Bigg[ 1 + 
                    \frac{\pi R\Lambda}{z} \left\{ 
                    \left( \frac{{\mu_1}}{\Lambda} \right)^2 
                  + \left( \frac{{\mu_2}}{\Lambda} \right)^2 
                    \right\} 
                    \coth (\pi z R\Lambda) 
                  + \frac{\pi^2}{2} (R\Lambda)^2 
                    \left(\frac{{\mu_1}}{\Lambda}\right)^2 
                    \left(\frac{{\mu_2}}{\Lambda}\right)^2  
                    \Bigg]. \nonumber \\ &&
\end{eqnarray} 

\subsection{Behavior of the Vacuum $\left< \mu_i \right>$}
\begin{figure}[htbp]
\centerline{\epsfig{figure=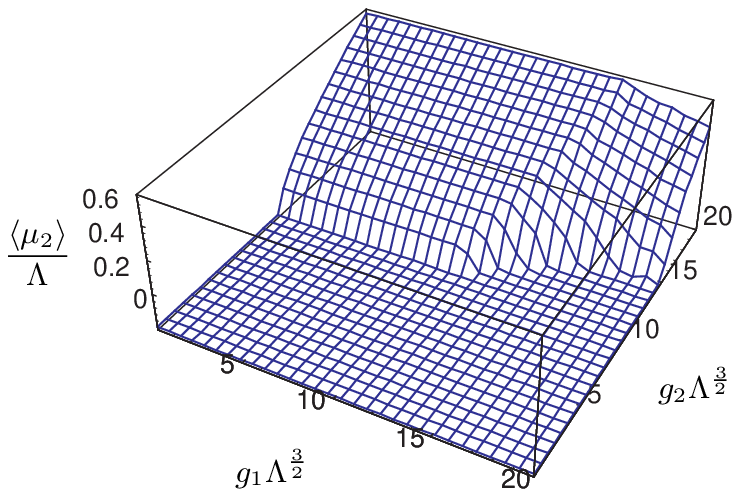,width=0.75\linewidth}}
\caption{$\left< \mu_2 \right>$ as a function of $g_1,g_2$.}
\label{3D}
\end{figure}

\begin{figure}[htbp]
  \centerline{\epsfig{figure=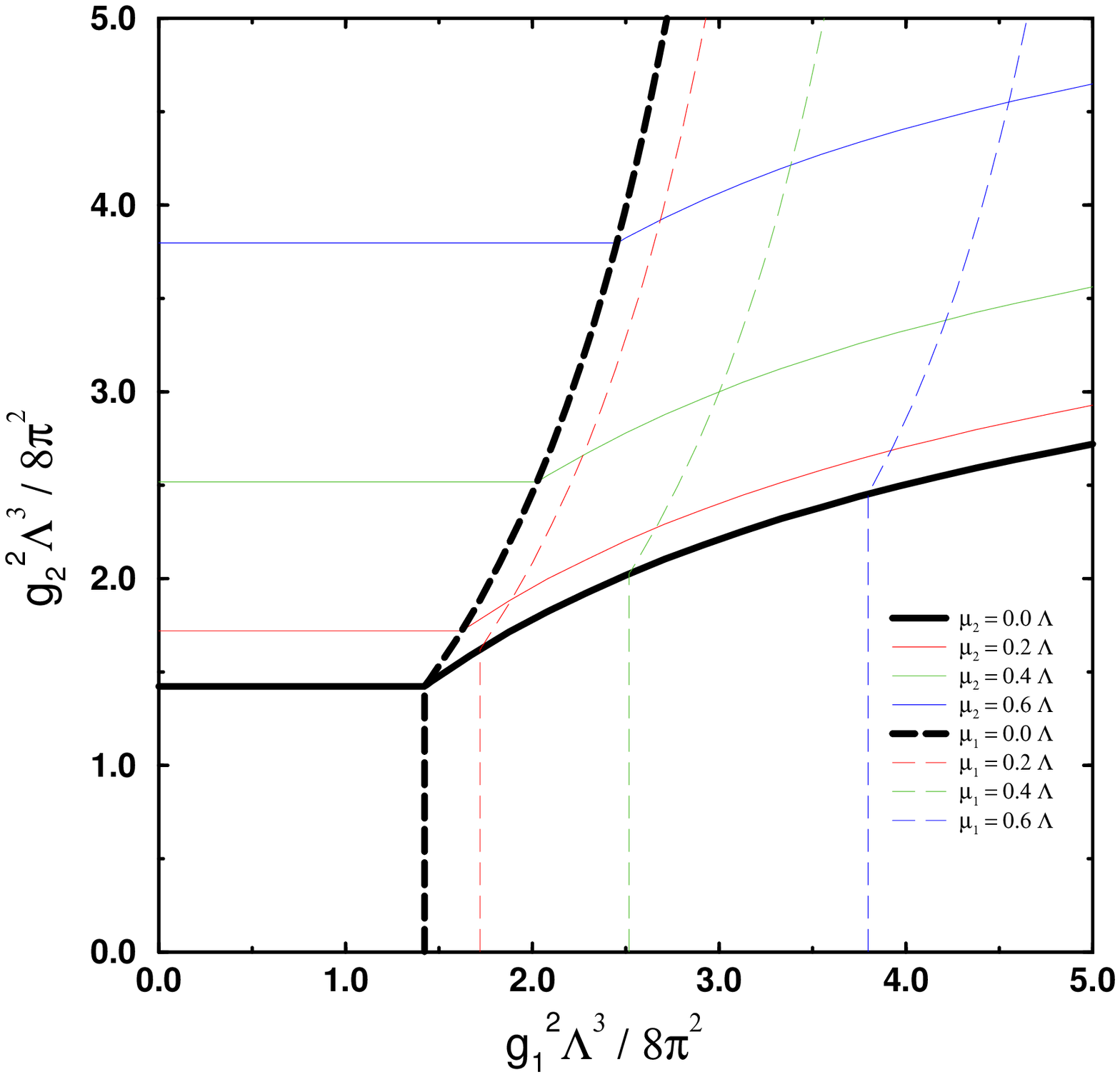,width=0.75\linewidth}}
\caption{The phase structure of the vacuum.}
\label{phase}
\end{figure}

\noindent
The behavior of the vacuum is determined by solving the gap equations 
($i,j=1,2$;\ $i\ne j$):
\begin{eqnarray}
\lefteqn{\frac{\partial \bar{V}}{\partial (\mu_i/\Lambda)}}
\hspace{0cm} \nonumber \\
 &=&  R\Lambda \frac{\mu_i}{\Lambda} \, 
      \Bigg\{ \frac{4\pi}{g_i^2 \Lambda^3} - \frac{1}{2\pi}\,
      \int_0^1 dz \,z^3 
         \frac{\pi R\Lambda \mu_j^2 \tanh(\pi z R\Lambda)+2}
         {z \left\{ 1+\frac{\pi^2}{2}(R\Lambda)^2 
                    \left( \frac{\mu_1}{\Lambda} \right) 
                    \left( \frac{\mu_2}{\Lambda} \right) \right\} 
          \tanh(\pi z R\Lambda) + \pi R\Lambda 
            \left\{ \left( \frac{\mu_1}{\Lambda} \right) + 
                    \left( \frac{\mu_2}{\Lambda} \right) \right\}}
      \Bigg\} \nonumber \\
 &=&   0.
\end{eqnarray}
The system has a second order phase transition and 
$\left< \mu_2 \right>$ as a function of $g_1$ and $g_2$ is shown in the Fig. 
\ref{3D} for $R\Lambda=1$.
If $g_i \gtrsim g_{i,{\rm critical}}$, we see $\left< \mu_i \right> \gtrsim 0$.
The phase strucutre of this system is summerized in the Fig. \ref{phase}.

\subsection{Effective Theory on $y=\frac{1}{2}$ brane}
\noindent
Now we realize that $\mu_1$ and $\mu_2$ can have non-zero vacuum expectation 
value and Lagrangian (\ref{4LagRS}) has mixing term 
$\bar{\psi}_RL_i + {\rm h.c.}$.
The physics on the $i=2$ brane is examined by 
integrating out the invisible field $L_1$. Setting 
$\sqrt{2}\psi^{(n)}_L \equiv N^{(n)}-M^{(n)}$ and 
$\sqrt{2}\psi^{(n)}_R \equiv N^{(n)}+M^{(n)}$\ ($n\ne 0$), 
the effective Lagrangian on the $y=\frac{1}{2}$ brane is obtained as follows: 
\begin{eqnarray}
{\cal L\/}^{(4)}_{\rm eff}
  =  \bar{\Theta}_2 [\, i\partial\!\!\!/ + M_2 + |\mu_1|^2 P \,] \Theta_2,
\label{eq:eff4Lag}
\end{eqnarray}
where 
$\Theta_2^T 
\equiv  \left( L_2 \, , \, \psi^{(0)}_R \, , \, N^{(1)} \, , \,
           M^{(1)} \, , \, N^{(2)} \, , \, M^{(2)} 
           \cdots \right)$ and 
\begin{eqnarray}
  M_2  \equiv  \left(
       \matrix{
       0 & \mu_2^{*} & -\mu_2^{*} & -\mu_2^{*} & \mu_2^{*} & \mu_2^{*} & \cdots \cr
       \mu_2 & 0 & 0 & 0 & 0 & 0 & \cdots \cr
      -\mu_2 & 0 & \frac{1}{R} & 0 & 0 & 0 & \cdots   \cr
      -\mu_2 & 0 & 0 & - \frac{1}{R} & 0 & 0 & \cdots   \cr
       \mu_2 & 0 & 0 & 0 & \frac{2}{R} & 0 & \cdots   \cr
       \mu_2 & 0 & 0 & 0 & 0 & - \frac{2}{R} & \cdots   \cr
       \vdots & \vdots & \vdots & \vdots & \vdots & \vdots & \ddots  
       }
               \right), \\
{\large P}  \equiv      \left(
         \matrix{
            0 & 0 & 0 & 0 & 0 & 0 & 
            \cdots \cr
            0 & (i\partial\!\!\!/)^{-1} & (i\partial\!\!\!/)^{-1} & 
            (i\partial\!\!\!/)^{-1} & (i\partial\!\!\!/)^{-1} & 
            (i\partial\!\!\!/)^{-1} & 
            \cdots \cr 
            0 & (i\partial\!\!\!/)^{-1} & (i\partial\!\!\!/)^{-1} & 
            (i\partial\!\!\!/)^{-1} & (i\partial\!\!\!/)^{-1} & 
            (i\partial\!\!\!/)^{-1} & 
            \cdots   \cr
            0 & (i\partial\!\!\!/)^{-1} & (i\partial\!\!\!/)^{-1} & 
            (i\partial\!\!\!/)^{-1} & (i\partial\!\!\!/)^{-1} & 
            (i\partial\!\!\!/)^{-1} & 
            \cdots   \cr
            0 & (i\partial\!\!\!/)^{-1} & (i\partial\!\!\!/)^{-1} & 
            (i\partial\!\!\!/)^{-1} & (i\partial\!\!\!/)^{-1} & 
            (i\partial\!\!\!/)^{-1} & \cdots   \cr
            0 & (i\partial\!\!\!/)^{-1} & (i\partial\!\!\!/)^{-1} & 
            (i\partial\!\!\!/)^{-1} & (i\partial\!\!\!/)^{-1} & 
            (i\partial\!\!\!/)^{-1} & \cdots   \cr          
            \vdots & \vdots & \vdots & \vdots & \vdots & \vdots & 
            \ddots  
         }
        \right).
\end{eqnarray}  
We have seen in the previous subsection that the dynamical fermion mass 
generation with $\psi$ is of the second order phase transition, and 
we find the parameter region of $(g_1,g_2)$ for 
$\left< \mu_2 \right> \lesssim \Lambda=1/R \sim {\rm TeV}$, 
$\left< \mu_1 \right> =0 $ for the cut-off $\Lambda=1/R$. 
Thus in Eq. (\ref{eq:eff4Lag}) we have a posibility to get a light Dirac 
fermion on the $i=2$ brane.

\section{Conclusion}
\noindent
We have found within our model (\ref{eq:5Lag}) 
that the dynamically generated fermion 
mass is much smaller than the Planck scale because of the presence of 
the mass scale $1/R \sim$ TeV generated by the Randall-Sundrum warp factor.
Furthermore, in spite of the presence of the 
mass scale $1/R \sim$ TeV in the theory the fermion masses on the 4-dimensional 
brane can be made smaller than this scale as a consequence of the interaction 
among the bulk and brane fermions: the mixing of the brane fermions with 
the bulk fermions does not lead to the lightest fermion masses 
of order $1/R$ and the dynamically generated fermion masses 
are not of order $1/R$ but of order $\left< \mu_2 \right>$ because of the same 
mechanism as Eq. (\ref{Mass}).
This result is obtained because the dynamical fermion masses generated 
under the second-order phase transition are small irrespectively of $1/R$ 
near the critical radius. In our model the possibility of having low mass 
fermions resulted from the dynamical origin. This mechanism is quite 
different from the ones in other approaches in which low mass fermions 
are expected to show up as a result of the kinematical 
origins.\cite{Ark2,Die,Moh,Das,Yos}

Our outlook is summarized as in the following four items. 
The first item is whether 
the K.-K. gauge boson (or graviton) exchange 
induce suitable effective four-fermion interactions, 
the second is what the physics on the visible ($i=2$) brane with 
$\left<\mu_1\right>\ne 0$ is, 
the third is how we stabilize the radius $b_0 = 24\pi k^{-1}$ 
with the pressure of the bulk fermion $\psi$,\cite{Gun,Abe,Gol}
and the last is the extension of our model to electroweak 
symmetry breaking.\cite{Dob,Chen,ArkPRD,Has,Riu}
\footnote{In the Ref.\cite{Riu}, the 
dynamical generation of fermion masses as well as electroweak symmetry 
breaking in the Randall-Sundrum scenario is studied. 
This paper also presents a scenario which can explain the observed hierarchy 
of fermion masses.}

\section*{Acknowledgments}
The authors would like to thank Hironori Miguchi, Koichi Yoshioka (Kyoto U.), 
Masahiro Yamaguchi (Tohoku U.) and Hiroaki Nakano (Niigata U.) 
for fruitful discussions and correspondences.
They are also indebted to Tak Morozumi for useful comments. 
The present work is supported financially by 
the Monbusho Grant, Grant-in-Aid for Scientific Research (C) with contract 
number 11640280.

\begin{thebibliography}{000}
\bibitem{KK}
Th. Kaluza, Sitzungsber. d. Preuss. Akad. d. Wiss. p.966 (1921);
O. Klein, Zeitsch. f. Phys. \textbf{37}, 895 (1926).
\bibitem{Ant}
I.~Antoniadis,
Phys.\ Lett.\ B {\bf 246}, 377 (1990);
I.~Antoniadis and K.~Benakli, 
Phys.\ Lett.\ B {\bf 326}, 69 (1994).
\bibitem{Ark1}
N. Arkani-Hamed, S. Dimopoulos and G. Dvali, Phys. Lett. \textbf{B429}, 263 (1998);
Phys. Rev. \textbf{D59}, 086004 (1999);
I. Antoniadis, N. Arkani-Hamed, S. Dimopoulos and G. Dvali, 
Phys.\ Lett.\ B {\bf 436}, 257 (1998).
\bibitem{Han}
T. Han, J. D. Lykken and R-J. Zhang, Phys. Rev. \textbf{D59}, 105006 (1999).
\bibitem{Ran}
L. Randall and R. Sundrum, Phys. Rev. Lett. {\bf 83}, 3370 (1999).
\bibitem{Cha}
S. Chang, J. Hisano, H. Nakano, N. Okada and M. Yamaguchi, 
Phys. Rev. {\bf D62}, 084025 (2000).
\bibitem{Dob}
B. A. Dobrescu, Phys. Lett. {\bf B461}, 99 (1999).
\bibitem{Chen}
H-C. Cheng, B. A. Dobrescu and C. T. Hill, Nucl. Phys. B {\bf 589}, 249 (2000);
A. B. Kobakhidze, hep-ph/9904203.
\bibitem{Abe1}
H. Abe, H. Miguchi and T. Muta, Mod. Phys. Lett. {\bf A15}, 445 (2000).
\bibitem{Ark2}
N. Arkani-Hamed, S. Dimopoulos, G. Dvali and J. March-Russel, hep-ph/9811448.
\bibitem{Die}
K. R. Dienes, E. Dudas and T. Gherghetta, Nucl. Phys. \textbf{B557}, 25 (1999).
\bibitem{Moh}
R. N. Mohapatra, S. Nandi and A. Perez-Lorenzana, Phys. Lett. \textbf{B466}, 115 (1999).
\bibitem{Das}
A. Das, and C. W. Kong, Phys. Lett. {\bf B470}, 149 (1999).
\bibitem{Yos}
K. Yoshioka, Mod. Phys. Lett. {\bf A15}, 29 (2000).
\bibitem{Gun}
S. A. Gundersen and F. Ravndal, Ann. Phys. \textbf{182}, 90 (1988).
\bibitem{Abe}
H. Abe, J. Hashida, T. Muta and A. Purwanto, Mod. Phys. Lett. \textbf{A14}, 1033 (1999).
\bibitem{Gol}
W. D. Goldberger and M. B. Wise, Phys. Rev. Lett. {\bf 83}, 4922 (1999).
\bibitem{ArkPRD}
N. Arkani-Hamed, H.-C. Cheng, B. A. Dobrescu and L. J. Hall, 
Phys. Rev. {\bf D62}, 096006 (2000).
\bibitem{Has}
M. Hashimoto, M. Tanabashi and K. Yamawaki, hep-ph/0010260
\bibitem{Riu}
N. Rius and V. Sanz, hep-ph/0103086.
\end{thebibliography}
\end{document}